\documentclass[a4paper, 11pt]{article}
\usepackage[]{graphicx}
\usepackage[dvipsnames,table]{xcolor}
\usepackage{alltt}
\PassOptionsToPackage{hyphens}{url}\usepackage{hyperref}
\usepackage{amsmath, amssymb}
\usepackage{doi} 
\usepackage[round]{natbib} 
\usepackage{multirow}
\usepackage{bbm}
\usepackage{algorithm}
\usepackage{algpseudocode}
\usepackage{microtype}
\usepackage{booktabs} 
\usepackage[title]{appendix} 
\usepackage{nameref} 
\usepackage[dvipsnames,table]{xcolor}
\usepackage[onehalfspacing]{setspace} 
\usepackage[labelfont=bf,font=small]{caption} 
\usepackage{newtxtext,newtxmath}
\usepackage{sectsty} 
\allsectionsfont{\sffamily} 
\usepackage[labelfont={bf,sf},font=small]{caption} 
\usepackage{orcidlink} 
\usepackage{fancyhdr}
\usepackage{geometry}
\makeatletter
\def\maxwidth{ %
  \ifdim\Gin@nat@width>\linewidth
    \linewidth
  \else
    \Gin@nat@width
  \fi
}
\makeatother

\usepackage{framed}
\makeatletter
 {\par\unskip\endMakeFramed%
 \at@end@of@kframe}
\makeatother

\definecolor{unipd_red}{HTML}{b40908}

\newcommand\mail{roberto.macridemartino@phd.unipd.it}
\title{\vspace{-2em}
  \textbf{\textbf{Alternative ranking measures to predict international football results}}
}
\author{
   \textbf{Roberto Macrì Demartino}\thanks{Corresponding author e-mail: \href{mailto:\mail}{\texttt{\mail}}} \textsuperscript{ $a$}   \orcidlink{0000-0002-5296-6566}, \textbf{Leonardo Egidi}\textsuperscript{$b$} \orcidlink{0000-0003-3211-905X}, and \textbf{Nicola Torelli}\textsuperscript{$b$}  \orcidlink{0000-0001-9523-5336}  \\
  \small\textsuperscript{$a$} Department of Statistical Sciences, University of Padova, Via C. Battisti 241, Padova, 35121, Italy. \\
  \small\textsuperscript{$b$} Department of Economics, Business, Mathematics and Statistics ``Bruno de Finetti", University of Trieste,\\
  \small Via A. Valerio 4/1, Trieste, 34127, Italy \vspace{2em} 
}
\date{}

\IfFileExists{upquote.sty}{\usepackage{upquote}}{}

\usepackage{hyperref}
\hypersetup{
  bookmarksopen=true,
  breaklinks=true,
  colorlinks=true,
  linkcolor=blue,
  anchorcolor=black,
  citecolor=blue,
  urlcolor=black,
}

\begin{document}

\maketitle

\begin{abstract}
Over the last few years, there has been a growing interest in the prediction and modelling of competitive sports outcomes, with particular emphasis placed on this area by the Bayesian statistics and machine learning communities. In this paper, we have carried out a comparative evaluation of statistical and machine learning models to assess their predictive performance for the 2022 FIFA World Cup and the 2023 CAF Africa Cup of Nations by evaluating alternative summaries of past performances related to the involved teams. More specifically, we consider the Bayesian Bradley-Terry-Davidson model, which is a widely used statistical framework for ranking items based on paired comparisons that have been applied successfully in various domains, including football. The analysis was performed including in some canonical goal-based models both the Bradley-Terry-Davidson derived ranking and the widely recognized Coca-Cola FIFA ranking commonly adopted by football fans and amateurs.
  \\[1ex]
\end{abstract}

\section{Introduction}\label{sec1}
The application of statistical and machine learning models in forecasting international football competitions, such as the FIFA World Cup or UEFA Champions League, has always attracted the interest of several analysts.

From a statistical perspective, the outcome of a football match may be predicted using two different approaches. The result-based approach \citep[among others]{koning2000, Carpita2019}, which directly predicts the match result, explicitly modelling the so-called three-way process – home win, draw, or away win – using a logistic or multinomial regression model. Alternatively, the goal-based approach models the goals scored and conceded in each match by modelling count variables, typically by using Poisson regressions, and then determines the exact match result by comparing these scores. It is worth noting that once a goal-based model has been estimated, it becomes possible to derive the three-way process by simply aggregating the estimated probabilities. However, for a more comprehensive comparison of goal-based and result-based statistical methods, see \citet{egidi_torelli_2021}.

This paper investigates the potential improvement in predictive performance achieved by some statistical goal-based methods and machine learning result-based algorithms when an appropriate measure of the teams' relative strength is used as an additional predictor. To assess it, we analyze the outcomes of the 2022 FIFA World Cup in Qatar and the 2023 Africa Cup in Ivory Coast. We considered a Bayesian Bradley-Terry-Davidson derived ranking system, using the posterior median of the log-strength parameters as a novel predictor. Subsequently, we compare the predictive performances of this approach with those obtained using the well-established FIFA ranking -- in terms of FIFA ranking points -- to determine which provides more reliable forecasts.

In the goal-based approach, each game involves evaluating the goal counts for each team. Furthermore, the expected number of goals is determined by team attributes such as offensive and defensive abilities, and home advantage when applicable.
Under some basic assumptions, the team-specific Poisson distributions are considered independent, resulting in a double Poisson model \citep[among others]{maher1982, Baio_Blangiardo2010, groll_abedieh_2013, egidi_etal2018}. However, these Poisson-based approaches can be generalized in different ways to capture the dependence between scores. For instance, \citet{dixon_coles1997} extended the work of \citet{maher1982} by allowing (a slightly negative) correlation between scores and incorporating a dependence parameter within their model to account for it. Furthermore, the bivariate Poisson model, designed to account for positive goal dependencies, was developed by \citet{karlis_ntzoufras2003} within a frequentist framework and by \citet{ntzoufras2011bayesian} from a Bayesian perspective. A key limitation of previous models lies in their assumption of invariant team-specific parameters, implying that team performance remains constant over time based on their offensive and defensive abilities. However, it is recognised that team performance is inherently dynamic, fluctuating over years and possibly within seasons. \citet{Rue_2000} proposed a dynamic extension for the double Poisson model on continuous time, while \citet{owen_2011} proposed a discrete time random walk approach for both offensive and defensive parameters. In addition, \citet{Koopman_2015} further extended the bivariate Poisson model into a state-space framework, allowing team abilities to vary according to a state vector.

A fundamentally different modelling approach has emerged due to the availability of large volumes of data, which has led to the development of a range of machine learning result-based techniques. These include artificial neural networks (ANNs), multivariate adaptive regression splines (MARS) \citep{Friedman_1991}, and ensemble learning methods such as random forests \citep{breiman2001random}. Notably, the predictive performance of several random forests configurations has been examined and evaluated in the context of international football matches \citep[among others]{Schauberger_Groll2018, groll2019, Groll2021}.

While the aforementioned methodologies provide robust frameworks for modelling football match outcomes, their predictive performance may be improved by incorporating additional historical information.
Typically, one approach to potentially improve predictive accuracy is to integrate established rankings, such as the FIFA ranking, as additional model covariates. Specifically, these covariates are based on the quantitative measures of team strength from which the overall ranking is then derived.
Notably, the algorithm underlying the FIFA ranking system has undergone a significant revision since 2018. The new algorithm takes into account not only the outcome of the single matches but also the strength level of teams before each match. For further details see \citet{szczecinski2022fifa}. The resulting algorithm offers benefits in terms of both simplicity and relative transparency. However, an alternative and particularly interesting methodology for obtaining a ranking, based on pairwise comparisons between teams, is the Bradley-Terry model \citep{bradley_Terry1952}. The model assigns a strength parameter to each team, and the odds of winning a match are determined by the ratio of these parameters. The estimated strengths can be employed to construct a rating system that reflects the ranking of teams based on the outcome of matches and the competitive interactions between teams.

In order to encompass a range of limitations, a number of extensions and generalizations to the original Bradley-Terry model have been developed over the years. For instance, \citet{rao_kupper_1967} and \citet{davidson1970} extended the model applicability to scenarios in which a draw is a possible outcome by including a novel parameter which affects the probability of a tie in a match. \citet{Springall_1973} proposed a generalization of the Bradley-Terry model with team-dependent linear predictors. \citet{Davidson_Solomon_1973} proposed a Bayesian version of the Bradley-Terry model, using a family of conjugate prior distributions to compute the posterior distribution of the log-strength parameters. Moreover, \citet{leonard1977} suggested a more flexible Bayesian approach using non-conjugated multivariate Gaussian prior distributions for the log-strengths. Since these early works, there have been numerous contributions that have further extended the Bayesian framework for paired comparison models  \citep[among others]{CHEN1984, Caron_Doucet_2012, whelan2017prior, OSEI2022, Wainer_2023}. The Bradley-Terry model can be further extended to account for situations where the order of comparisons can influence the outcome. A classic example is the home-field advantage in football, where the team playing at home may have a psychological or logistical advantage compared to the visiting team. Specifically, \citet{Beaver_Gokhale_1975} and \citet{davidson_beaver_1977} introduced additive and multiplicative order effects, respectively.  Additionally, several extensions have been developed to model the dynamic variation of strengths over time. \citep[among others]{Fahrmeir_Tutz_1994, Glickman_1999, Glickman_2001,Cattelan_et_al_2012, tian2023spectral}.

Our work focuses on extending the well-known goal-based and result-based protocols by introducing alternative ranking measures for international football matches that could serve as a valuable computational routine for practitioners. The rest of the paper is organized as follows. Section \ref{sec2} presents the theoretical framework, introducing the standard Bradley-Terry model and its Davidson extension for handling draws, followed by a discussion of the Bayesian approach. Furthermore, Section \ref{sect_Stat_ML_foot} describes the statistical goal-based methods and the machine learning result-based algorithms used in this study. 
In Section \ref{sec3}, we evaluate the application of these methodologies on the data from both the last World Cup and Africa Cup of Nations. Finally, Section \ref{sec4} provides concluding remarks, outlining the limitations, advantages, and potential future research directions.

\section{The Bradley-Terry Model}\label{sec2}
The Bradley-Terry model \citep{bradley_Terry1952} is one of the most popular modelling techniques in a pairwise comparison context for ranking players or teams. The model assumes that each team $T_k$, with $k = 1, \dots, N_T $,
is characterized by a latent parameter, $\alpha_k > 0$, representing its intrinsic strength, 
The outcome of any given comparison is modelled as an independent Bernoulli random variable, where the probability of each outcome is a function of the strengths of the teams involved. Specifically, for a match between team $T_i$ and team $T_j$, with $i\neq j = 1, \dots, N_T $ , the probability that $T_i$ defeats $T_j$ in the \textit{n}-th match, with $n = 1, \dots, N$, is 
\begin{equation}
    p_{ij}^W = \mathbb{P}(T_{i}\; \text{defeats} \; T_{j}) = \dfrac{\alpha_i}{\alpha_i+\alpha_j}, 
\label{basic_BT}
\end{equation}
where $\alpha_{i}$ and $\alpha_{j}$ are the strength parameters of the teams involved in the match. These parameters are invariant to a multiplicative constant. Therefore, parameter identifiability is obtained by imposing a constraint such as $\sum_{k=1}^{N_T} \alpha_k=1$. 
Furthermore, the final ranking of the teams can be determined by sorting their respective strength parameters $\alpha_k$.

The model in \eqref{basic_BT} is commonly reparameterized by the logarithm of the strength parameters 
\begin{equation}
   p_{ij}^W = \dfrac{\exp(\psi_i)}{\exp(\psi_i)+\exp(\psi_j)},
\label{log_par_BT}
\end{equation}
where $\psi_i = \log(\alpha_i)$ and $\psi_j = \log(\alpha_j)$. Since the $\alpha$ values are invariant to multiplicative constants, the $\psi$ values are invariant to additive constants. Consequently, the parameters are identifiable if $\sum_{k=1}^{N_T} \psi_k=0$. This transformation offers several advantages. Notably, it enables the estimation of the log-strength parameter across an expanded parameter space, $\psi \in (-\infty, +\infty)$, providing greater flexibility for the application of a wide class of priors in the Bayesian setting. Furthermore, the logit transformation facilitates parameter estimation within the frequentist framework using generalized linear models (GLMs) \citep{Cattelan_2012}.
\subsection{Dealing with draws}
The standard Bradley-Terry model does not account for draws. Several alternatives have been proposed to address this limitation, including the assignment of draws as wins to both teams, the spreading of draws as half a win to each team, the non-consideration of draws as wins, and the random assignment of draws as wins to one of the two teams. However, none of these approaches directly incorporates the possibility of a tie into the model.

To address this, \citet{rao_kupper_1967} extended the Bradley-Terry model to accommodate draws by introducing an additional parameter $\eta$, and explicitly modelling its probability as follows
\begin{equation*}
  \label{eq:t}
  \begin{aligned}
    p_{ij}^W &= \dfrac{\alpha_i}{\alpha_i+\eta\alpha_j},\\        
    p_{ij}^D &=   \mathbb{P}(T_{i}\; \text{draw} \; T_{j}) = \dfrac{(\eta^2-1)\alpha_i\alpha_j}{(\alpha_i+\eta\alpha_j)(\eta\alpha_i+\alpha_j)}.
  \end{aligned}
\end{equation*}
If $\eta = 1$ then the Rao-Kupper model reduces to the standard Bradley-Terry model.
Furthermore, using the log-parametrization as in \eqref{log_par_BT}, the model is
\begin{equation*}
  \label{eq:t}
  \begin{aligned}
    p_{ij}^W &= \dfrac{\exp(\psi_i)}{\exp(\psi_i)+\exp(\gamma+\psi_j)},\\        
    p_{ij}^D &=  \dfrac{(\exp(2\gamma)-1)\exp(\psi_i+\psi_j)}{[\exp(\psi_i)+\exp(\gamma+\psi_j)][\exp(\gamma+\psi_i)+\exp(\psi_j)]},
  \end{aligned}
\end{equation*}
where $\gamma = \log(\eta)$. 

An alternative approach, which adheres to the ratio scale required by the so-called choice axiom \citep{luce1959individual}, was proposed by \citet{davidson1970}. As in \citet{rao_kupper_1967}, the Bradley-Terry-Davidson (BTD) model introduces an additional parameter that balances the probability of ties against the probability of not having ties and computes two different probabilities. The log-parametrization of the model is 
\begin{equation}
  \label{davidson_ties}
  \begin{aligned}
    p_{ij}^W &= \dfrac{\exp(\psi_i)}{\exp(\psi_i)+\exp(\psi_j)+\exp(\gamma+ (\psi_i+\psi_j)/2)},\\  
    p_{ij}^D &=  \dfrac{\exp(\gamma+(\psi_i+\psi_j)/2)}{\exp(\psi_i)+\exp(\psi_j)+\exp(\gamma+ (\psi_i+\psi_j)/2)},\\
    p_{ij}^L &= \dfrac{\exp(\psi_j)}{\exp(\psi_i)+\exp(\psi_j)+\exp(\gamma+ (\psi_i+\psi_j)/2)}.
  \end{aligned}
\end{equation}
Since the three-way process events are mutually exclusive, the following constraint is imposed
$ p_{ij}^W + p_{ij}^D + p_{ij}^L = 1.
$
It is worth noticing that if the draw parameter $\gamma$ increases towards $+\infty$ then the probability of a tie $p_{ij}^D$ approaches one. Conversely, if $\gamma$ decreases towards $-\infty$ then $p_{ij}^D$ approaches to zero. Finally, if $\gamma$ is equal to zero then $p_{ij}^W$, $p_{ij}^D$, and $p_{ij}^L$ depend solely on the strengths of the competing teams.

For the remainder of the paper, we will focus specifically on Davidson's proposal for dealing with draws.

\subsection{The Bayesian approach}
 In the frequentist paradigm, teams are ranked using maximum likelihood estimates (MLE) of the strength parameters \citep{Ford1957SolutionOA, Hunter_2004}. However, the Bayesian framework offers a different perspective by providing the posterior distribution of the strength parameters reflecting the inherent uncertainty in the ranking system. 
Here, we introduce the hierarchical Bayesian formulation of the Bradley-Terry model incorporating the Davidson extension for handling draws as in \eqref{davidson_ties}.

The Bayesian BTD model requires specifying prior distributions for both the team log-strength parameters and the draw parameter. Specifically, the prior placed on $\gamma$ reflects our initial belief about the impact of teams' strengths on tie outcomes \citep{IssaMattos2022}.
When defining priors within this framework, \citet{whelan2017prior} proposed a set of desirable properties specifically suited for ranking systems. These properties aim to construct priors that avoid introducing unfair advantages or disadvantages for any particular team. Ideally, the prior should maintain invariance when teams are swapped, should not be affected by switching the outcome of the match for any given comparison, removing teams from the competition should not alter the prior distribution, and the prior should be proper. Specifically, employing a multivariate Gaussian for the log-strengths \citep{leonard1977}, or identical independent Gaussian distributions for each log-strength parameter, satisfies all the four conditions.

 Based on this, let $w_{ij}$ represent the binary outcome where team $T_i$ defeats team $T_j$, and let $d_{ij}$ indicate the binary outcome of a draw between teams $T_i$ and $T_j$. Then, the hierarchical Bayesian BTD model is
\begin{equation}
  \label{Bayesian_BTD}
  \begin{aligned}
    w_{ij} \mid p_{ij}^W &\sim \mathrm{Bernoulli}(p_{ij}^W),\\ 
    d_{ij} \mid p_{ij}^D &\sim \mathrm{Bernoulli}(p_{ij}^D),\\ 
    \psi &\sim \mathrm{N}(\mu_\psi,\sigma^2_\psi),\\        
    \gamma &\sim \mathrm{N}(\mu_\gamma,\sigma^2_\gamma),
  \end{aligned}
\end{equation}
where $\mu_\psi$ and $\mu_\gamma$ are the mean for the team log-strength and draw parameters, and $\sigma^2_\psi$ and $\sigma^2_\gamma$ denote the corresponding variances. 

\section{Statistical models and machine learning algorithms} \label{sect_Stat_ML_foot}
This section describes the statistical models and machine learning algorithms employed to predict the outcomes of the considered competitions. Through a detailed analysis of goal-based models and result-based machine learning algorithms, this section aims to provide a comprehensive overview of the methodologies employed in the prediction of football matches, emphasizing their statistical foundations and practical implementations in sports analytics.
\subsection{Goal-based models}
\label{sect_Goal-based_models}
Goal-based models assume that the number of goals scored in a match by each team follows a discrete distribution, typically two independent Poisson or a bivariate Poisson accounting for positive correlation. Thus, for each match, we need to consider the pair of counts $(X_{in}, Y_{jn})$, for $i\neq j = 1, \dots, N_T $ and $n = 1,\dots, N$. The first count $X_{in}$ denotes the non-negative number of goals scored by the home team $T_i$ and the second count $Y_{jn}$ denotes the number of goals scored by the visiting team $T_j$, both in the \textit{n}-th match. A simple double Poisson model is 
\begin{equation}
\label{double_poisson}
\begin{aligned}
X_{in} \mid \lambda_{1n} & \sim \operatorname{Poisson}\left(\lambda_{1 n}\right) \\
Y_{jn} \mid \lambda_{2n} & \sim \operatorname{Poisson}\left(\lambda_{2n}\right) \\
\log \left(\lambda_{1 n}\right) & =\theta+\operatorname{att}_{h_n}+\operatorname{def}_{a_n}+\frac{\phi}{2} \omega_n, \\
\log \left(\lambda_{2 n}\right) & =\theta+\operatorname{att}_{a_n}+\operatorname{def}_{h_n}-\frac{\phi}{2} \omega_n,
\end{aligned}
\end{equation}
where $\lambda_{1n}$ and $\lambda_{2n}$ describe the expected number of goals for the home team and the away team, respectively. In particular, $\theta$ denotes a common baseline parameter, the parameters $\textit{att}$ and $\textit{def}$ represent the unknown attack and defense abilities for the home team $h_n$ and the away team $a_n$ in the \textit{n}-th match. 
Furthermore, $\omega_n = (rank\_points_{h_n}-rank\_points_{a_n})$ captures the difference in FIFA ranking points (BTD relative log-strengths) between the home and away teams in the  \textit{n}-th match. Finally, the parameter $\phi$ tries to correct for the ranking points difference occurring between two competing teams. A sum-to-zero constraint \citep{Baio_Blangiardo2010} is imposed on the attack and defense parameters to ensure model identifiability.

A key limitation of the double Poisson model lies in its assumption of conditional independence between the goals scored by competing teams. However, in interactive team sports like football, a degree of correlation between goal outcomes is likely due to on-field interactions. This correlation could reflect changes in playing style by one or both teams throughout the match.
To address this limitation and account for the positive dependence between goal counts, a bivariate Poisson model \citep{karlis_ntzoufras2003} for each pair of counts can be considered
\begin{equation}
\label{bivariate_poisson}
\begin{aligned}
\left(X_{in}, Y_{jn} \mid \lambda_{1 n}, \lambda_{2 n}, \lambda_{3 n}\right) & \sim \operatorname{BivPoisson}\left(\lambda_{1 n}, \lambda_{2 n}, \lambda_{3 n}\right) \\
\log \left(\lambda_{3 n}\right) & =\beta_0,
\end{aligned}
\end{equation}
where $\lambda_{1n}$ and $\lambda_{2n}$ are defined as in \eqref{double_poisson}, whereas the coefficient $\lambda_{3n}$ describes the dependence between the two random counts. Furthermore, all the other parameters have the same interpretation as in \eqref{double_poisson}. Notably, when $\lambda_{3n}=0$, the two components are independent, then the bivariate Poisson model reduces to a double Poisson model. We note that in \eqref{bivariate_poisson} we let the covariance $\lambda_{3n}$ to not depend on other predictors, thus we assume it is equal for each match $n$: However, one could assume an extended linear predictor with match-dependent covariates, as specified in \cite{karlis_ntzoufras2003}.

Poisson goal-based models may suffer from an underestimation of the number of draws, represented by the outcomes in the diagonal of the probability table. To address this issue, \citet{karlis_ntzoufras2009} introduced a zero-inflated model for favoring the draw outcome. The diagonal-inflated bivariate Poisson model is defined as follows
\begin{equation}
\label{diag_poisson}
    \mathbb{P}(X_n=x_n, Y_n=y_n)= \begin{cases}(1-p) \operatorname{BP}\left(\lambda_{1n}, \lambda_{2n}, \lambda_{3n}\right) & \text { if } x_n \neq y_n \\ (1-p) \operatorname{BP}\left(\lambda_{1n}, \lambda_{2n}, \lambda_{3n}\right)+p D(x_n, \xi) & \text { if } x_n=y_n\end{cases},
\end{equation}
where $D(x_n, \xi)$ is a discrete distribution with parameter vector $\xi$.

Following \citet{owen_2011} and \citet{egidi_etal2018}, we introduce a dynamic assumption regarding team-specific effects for the models presented in equations \eqref{double_poisson}, \eqref{bivariate_poisson} and \eqref{diag_poisson}. A first-order autoregressive model is adopted by centering the effect of seasonal time $\tau$ on the lagged effect in $\tau-1$, plus a fixed effect. This allows attack and defense parameters to vary across seasons, where a season corresponds to a year. Therefore, for each team $i$, where $i = 1, \ldots, N_T$, and each year $\tau$, where $\tau = 2,\ldots,\mathcal{T}$, the prior distributions for the attack and defense abilities are defined as follows
\begin{equation*}
 \begin{aligned}
& \operatorname{att}_{i, \tau} \sim \mathrm{N}\left(\operatorname{att}_{i, \tau-1}, \sigma^2_{\operatorname{att}}\right) \\
& \operatorname{def}_{i, \tau} \sim \mathrm{N}\left(\operatorname{def}_{i, \tau-1}, \sigma^2_{\operatorname{def}}\right).
\end{aligned}   
\end{equation*}
For the initial season $\tau=1$, the prior distributions are initialized as
\begin{equation*}
\begin{aligned}
& \operatorname{att}_{i, 1} \sim \mathrm{N}\left(\mu_{\operatorname{att}}, \sigma^2_{\operatorname{att}}\right) \\
& \operatorname{def}_{i, 1} \sim \mathrm{N}\left(\mu_{\operatorname{def}}, \sigma^2_{\operatorname{def}}\right),
\end{aligned}
\end{equation*}
where $\mu_{\operatorname{att}}$ and $\mu_{\operatorname{def}}$ are the mean for the initial attack and defense abilities, and $\sigma_{\operatorname{att}}$ and $\sigma_{\operatorname{def}}$ are their corresponding variances. As with the static models, the dynamic extension also imposes a sum-to-zero constraint on these random effects within each season for identifiability.

\subsection{Result-based algorithms}
\label{sect_result-based_models}
Random forests \citep{breiman2001random} are ensemble learning algorithms that combine the predictions of a large number of decision trees. These methods are typically constructed from a large number of classification trees grown on bootstrap samples drawn from the original dataset. Notably, the aggregation of multiple trees offers several advantages. The resulting predictions inherit the unbiasedness of individual trees while exhibiting reduced variance. Additionally, the trees within a random forest are grown independently. This independence helps to reduce the overall variance of the ensemble compared to a single tree. To achieve this goal, random forests typically incorporate two key randomization steps during the tree building process.
Furthermore, several studies have demonstrated the efficacy of random forests in predicting international football match outcomes. These contributions consistently report better performance compared to regression approaches \citep[among others]{Schauberger_Groll2018, groll2019, Groll2021}.

Artificial neural networks (ANNs) represent a class of complex computational models inspired by the interconnected structure of neurons in the human brain. These models excel at processing information and learning from data through a layered architecture. Each layer comprises interconnected nodes that apply weights and biases to process inputs. The learning process involves adjusting these weights and biases to optimize the network's performance. Specifically,
ANNs proved successful in predicting football match outcomes by considering historical data that include a wide range of information, such as team performance metrics, match results, and even individual player statistics \citep[among others]{Huang_2010, Hucaljuk2011PredictingFS, Danisik_2018}. 

Multivariate Adaptive Regression Splines (MARS) was first proposed by \citet{Friedman_1991} as an algorithm to model non-linear relationships, particularly those that are nearly additive or involve low-order interactions between variables. Essentially, the algorithmic procedure involves a piecewise linear regression model. This allows the slope of the regression
line to change from one interval to the other as the two knots are crossed. The selection of variables and knot locations is determined through a computationally efficient but intensive forward-backward search procedure. Notably, \citet{Abreu2013} applied a MARS algorithm to investigate the relationship between the number of goals scored and the final game statistics.

\subsection{Computational procedure}
In the statistical models and machine learning algorithms described in Sections \ref{sect_Goal-based_models} and \ref{sect_result-based_models} an additional predictor, determined by the difference in FIFA ranking points (BTD relative log-strengths) between the home team and the away team, is incorporated. Specifically, for the BTD relative log-strengths, this process involves initially fitting the Bayesian BTD model as described in Equation \eqref{Bayesian_BTD}. Subsequently, the posterior median for each team's log-strength parameter is computed. Finally, the difference in the posterior medians of the competing teams is included as an additional predictor in both the goal-based models and result-based algorithms.

 The computational steps for integrating the Bayesian BTD relative log-strengths into the statistical models and machine learning algorithms are summarized in Algorithm 1.

\begin{algorithm}
\caption{Bayesian BTD computational steps}
\begin{algorithmic}[1]
\State Fit the Bayesian BTD model as described in \eqref{Bayesian_BTD}.
\State For $k=1,\ldots,N_T$ compute the posterior median for each team's log-strength parameter $\psi_k$.
\State For $n=1,\ldots,N$ incorporate the difference in the posterior median of the competing teams' log-strengths, $\omega_n = \psi_{h_n}-\psi_{a_n}$, as an additional predictor in both the goal-based models and result-based algorithms for the \textit{n}-th match of the competition.
\end{algorithmic}
\end{algorithm}
 
\section{Applications}
\label{sec3}
The selection of the training and test sets is crucial and is likely to influence the predictions.
We address this by employing an iterative training approach. Our goal-based statistical models and result-based machine learning algorithms are trained on a continuously updated dataset encompassing international matches from $2018$ to $2023$. The matches vary from the FIFA World Cups through the UEFA Euro Championships to normal friendly matches. The data excludes Olympic Games and matches in which at least one of the teams was the national B-team or a U-23 lineup. This approach allows for the incorporation of recent results, potentially improving predictive performance for two distinct scenarios: the group stage and the knockout stage of these two tournaments.

We evaluate the predictive performance of three dynamic goal-based Poisson models, implemented using the \texttt{footBayes} package \citep{footbayes}, and three result-based machine learning techniques provided by the \texttt{caret} package \citep{caret}, both implemented in R, along the lines described in Section~\ref{sect_Stat_ML_foot}. 
Furthermore, we incorporate additional historical information from both the FIFA ranking -- through the FIFA ranking points -- and the Bayesian BTD derived ranking -- through its relative log-strengths -- using the \texttt{bpcs} R package \citep{issa_mattos_package}. 


To ensure a more comparable analysis, both FIFA ranking points and BTD relative log-strengths are normalized using the scaled median absolute deviation (MAD) normalization
\begin{equation*}
    x_{\mathrm{MAD}} = \dfrac{x-\mathbb{M}(x)}{\mathbb{M}\left(|x-\mathbb{M}(x)|\right)},
\end{equation*}
where $\mathbb{M}(\cdot)$ is the median. Furthermore, to assess the predictive performance of the models described in Section \ref{sect_Stat_ML_foot}, we employed the Brier score \citep{brier1950verification} as recommended by \citet{spiegelhalter2009one}. It is essentially a mean squared error for forecasts
where a lower score indicates greater model predictive accuracy. A common formulation is 
\begin{equation*}
    b = \dfrac{1}{N}\sum_{n=1}^N\sum_{r=1}^3(p_{rn}-\delta_{rn})^2,
\end{equation*}
where $p_{rn}$ represents the predicted probability of outcome $r$, with $r \in \left\{\text{win, draw, loss}\right\}$, for the \textit{n}-th match. Here, $\delta_{rn}$ denotes the Kronecker delta, which equals $1$ if the actual outcome of the \textit{n}-th match corresponds to $r$. 
The lower bound of the Brier score is 0, which occurs when the predicted probabilities are perfectly accurate. In the case of three categories, the upper bound of the Brier score is 2. This happens when the worst prediction is made by assigning a probability of 1 to an incorrect category and 0 to the correct category. This results in a squared difference of 1 for both the correct and the selected incorrect category. Thus, if two such errors are made, the sum is 2.
\subsection{2022 FIFA World Cup}
\label{subsec:fifa2022_application}
 The World Cup presents an interesting case study due to the diverse range of National teams with heterogeneous strengths. We specifically investigate the performance of teams in a structured environment, such as the group stage, and contrast it with the dynamic and high-pressure setting of the knockout stage, where single-elimination games can significantly impact teams' behaviour. This subsection describes how we employ both goal-based statistical models and machine learning algorithms to forecast match outcomes throughout the tournament, assessing their predictive performance by adding the historical information from the ranking systems outlined previously. Notably, we consider the FIFA ranking published just before the World Cup took place, available at \url{https://inside.fifa.com/fifa-world-ranking/men?dateId=id13869}.

\begin{figure}[htb!]
    \centering   \includegraphics[width=12cm, height=9.5cm]{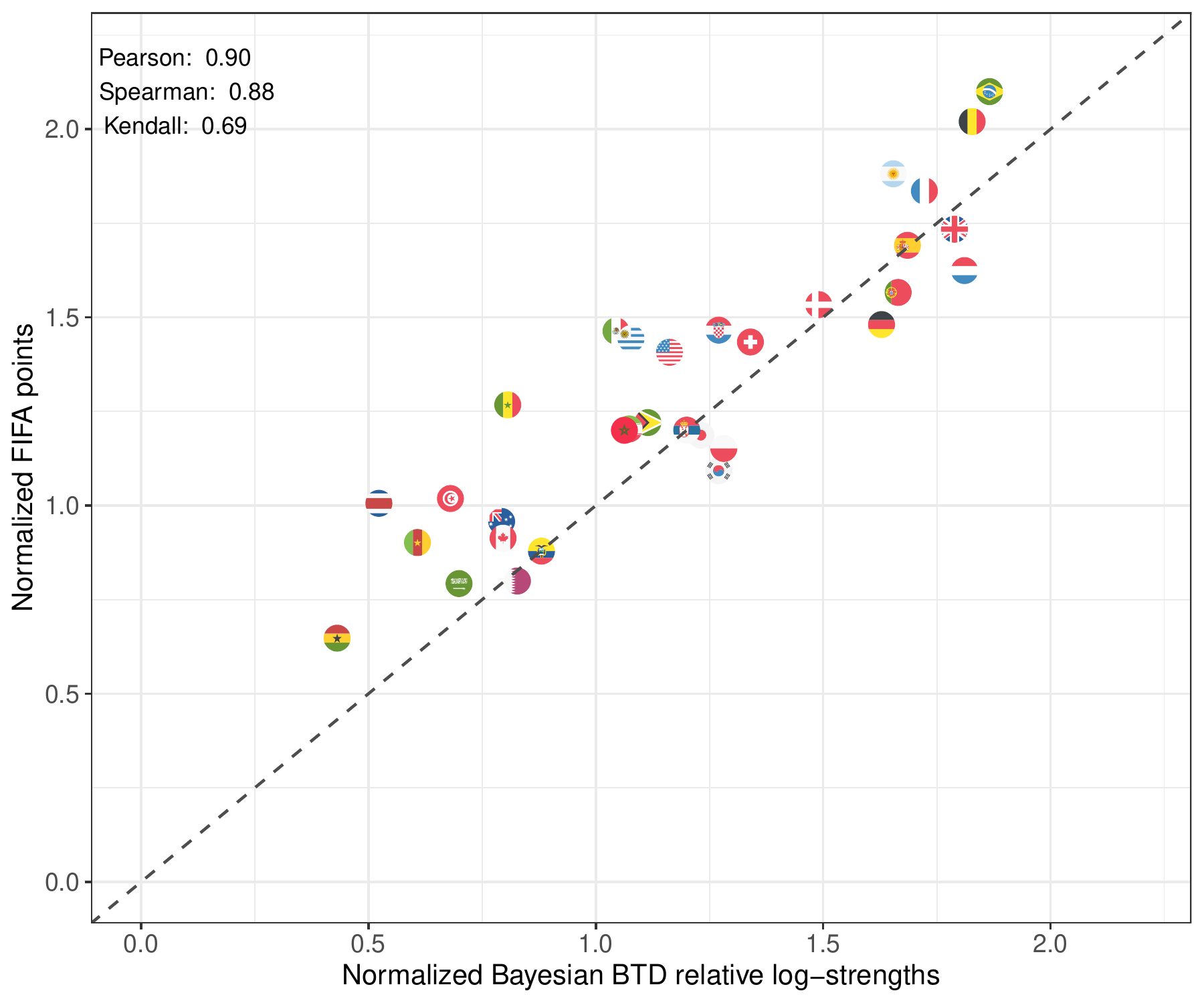}
        \caption{2022 FIFA World Cup. Scatterplot comparing the FIFA ranking points and the BTD relative log-strengths of teams in the World Cup.}
        \label{ranks_comp_WC}
\end{figure}

Figure \ref{ranks_comp_WC} presents scatterplot between the normalized Bayesian BTD relative log-strengths and the normalized FIFA points. The high value of the Pearson correlation coefficient ($\rho_p = 0.90$) suggests a positive linear relationship between these two variables. This is further corroborated by the Spearman correlation coefficient  ($\rho_s = 0.88$). Additionally, the Kendall coefficient confirms a substantial positive association ($\tau = 0.69$), albeit slightly weaker than the Pearson and Spearman correlations.

\begin{figure}[htb!]
    \centering   \includegraphics[width=14.5cm, height=14.5cm]{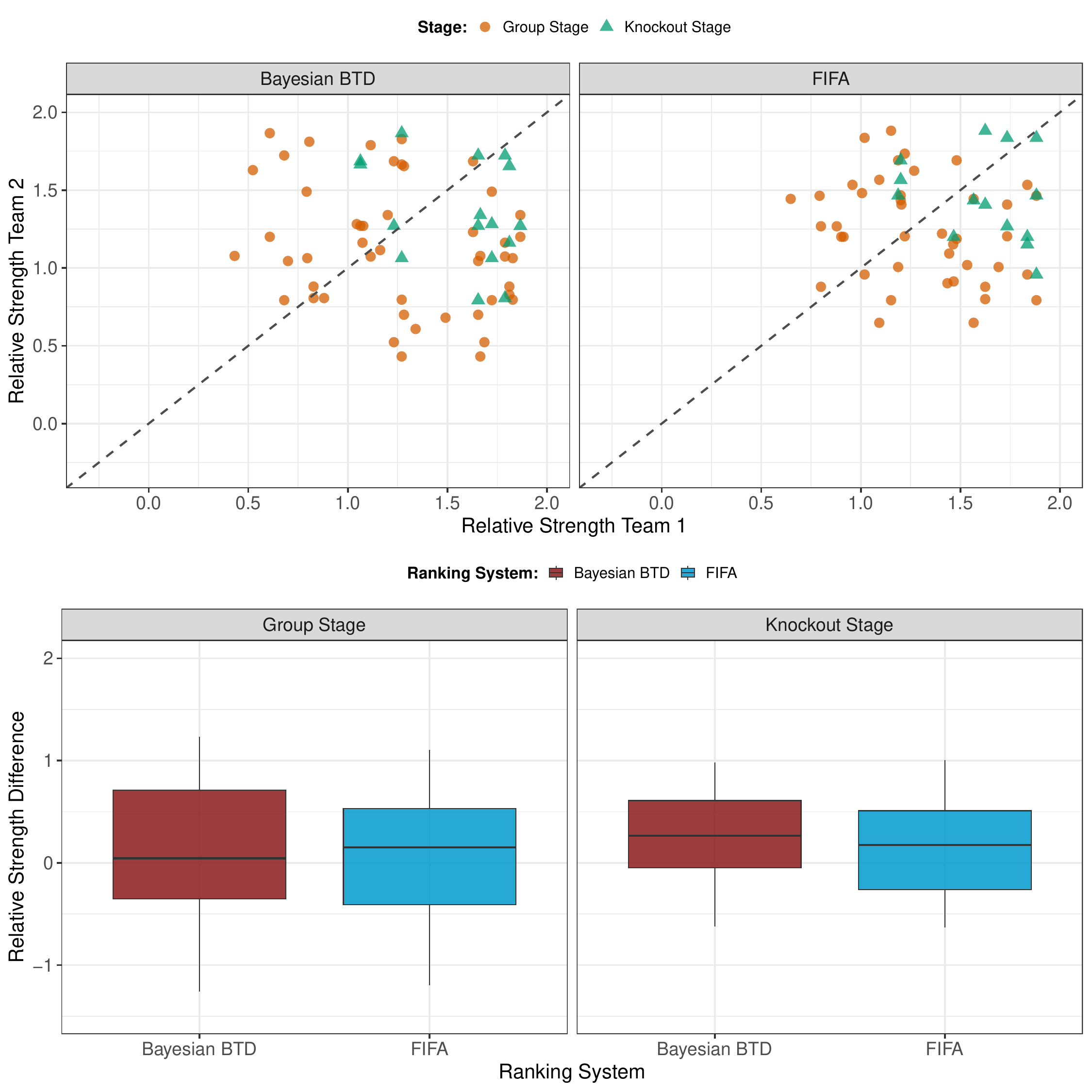}
        \caption{2022 FIFA World Cup. The top panel displays scatterplots comparing the normalized relative strengths of teams in both the group stage (orange dots) and knockout stage (green dots) under the two ranking systems. The dashed grey line represents the bisector. The bottom panel presents boxplots of the normalized relative strength differences for the FIFA ranking (blue) and the Bayesian BTD derived ranking (red) across the two World Cup stages.}
        \label{ranks_WC}
\end{figure}

The top panel of Figure \ref{ranks_WC} presents scatterplots comparing the relative strengths of competing teams in both the group and knockout stages under the two ranking systems. Points above the dashed line represent matches where the ``away" team had higher relative strength than the ``home" team, while points below the line indicate the opposite - we note that the terms `home' and `away' do not mean anything relevant in an international competition, where there are just one or two hosting teams. As expected, both the ranking systems reveal greater variability in team relative strengths during the group stage compared to the knockout stage. This is because teams in the group stage are typically more heterogeneous in terms of strength. As the tournament advances to the knockout stage, the remaining teams become increasingly similar in ability, leading to less variation in relative strength. This pattern is even more evident in the bottom panel of Figure \ref{ranks_WC}, which displays boxplots of relative strength differences between competing teams for each ranking system across the two stages. The boxplots further reveal that the Bayesian BTD ranking system exhibits more variability than the FIFA ranking during the group stage. In contrast, during the knockout stage, relative strength differences variability under both ranking systems are more concentrated, indicating a closer similarity in team abilities.

The main appeal of these models lies in their ability to predict the outcome of football matches. Table \ref{brier_WC} illustrates the predictive accuracy of dynamic Poisson models and machine learning algorithms, measured by the Brier Score, across both group and knockout stages. The goal-based statistical models perform slightly better than the result-based methods during the group stage. In contrast, the machine learning approaches show better predictive accuracy in the knockout stage, where results may be less predictable. A similar trend is seen with ranking systems. While the FIFA ranking system shows marginally better predictive accuracy in the group stage, the Bayesian BTD ranking demonstrates overall better performance during the knockout stage. This suggests that the Bayesian BTD relative log-strengths are particularly apt at predicting outcomes when teams have comparable abilities, which is often the case in the latter stages of the competition.
\begin{table}[t!]
\centering
\begin{tabular}{lcccc}
\toprule
\multirow{2}{*}{\textbf{Model}} & \multicolumn{2}{c}{\textbf{Group Stage}} & \multicolumn{2}{c}{\textbf{Knockout Stage}} \\ 
\cmidrule(lr){2-3} \cmidrule(lr){4-5} 
 & FIFA & BTD & FIFA & BTD \\ 
\midrule
Diag. Infl. & 0.620 & 0.629 & 0.530 & 0.510 \\ 
Biv. Pois. & 0.617 & 0.618 & 0.546 & 0.535 \\ 
Double Pois. & 0.622 & 0.623 & 0.543 & 0.527 \\ 
MARS & 0.640 & 0.660 & 0.486 & 0.503 \\ 
ANN & 0.627 & 0.660 & 0.465 & 0.471 \\ 
Random Forest & 0.713 & 0.745 & 0.493 & 0.461 \\ 
\bottomrule
\end{tabular}
\caption{2022 FIFA World Cup. Brier score for the FIFA ranking and the Bayesian BTD derived ranking across the two World Cup stages.}
\label{brier_WC}
\end{table}
\subsection{2023 CAF Africa Cup of Nations}
In this section, we fit the considered statistical models and machine learning algorithms to the data from the most recent CAF Africa Cup of Nations (AFCON) tournament held in Ivory Coast. Notably, the AFCON competition differs from the FIFA World Cup in that the participating teams tend to be more similar in terms of overall strength even during the group stage, making it a compelling case for analyzing the effectiveness of the Bayesian BTD ranking system. In order to conduct this analysis, we use as training set the data from matches played throughout 2018 to the end of 2023. Furthermore, the Bayesian BTD model was executed within this same period to generate team relative log-strengths. In addition, the FIFA ranking employed corresponds to those published on December 21st, available at \url{https://inside.fifa.com/fifa-world-ranking/men?dateId=id14233}.

\begin{figure}[tb!]
\centering
\includegraphics[width=12cm, height=9.5cm]{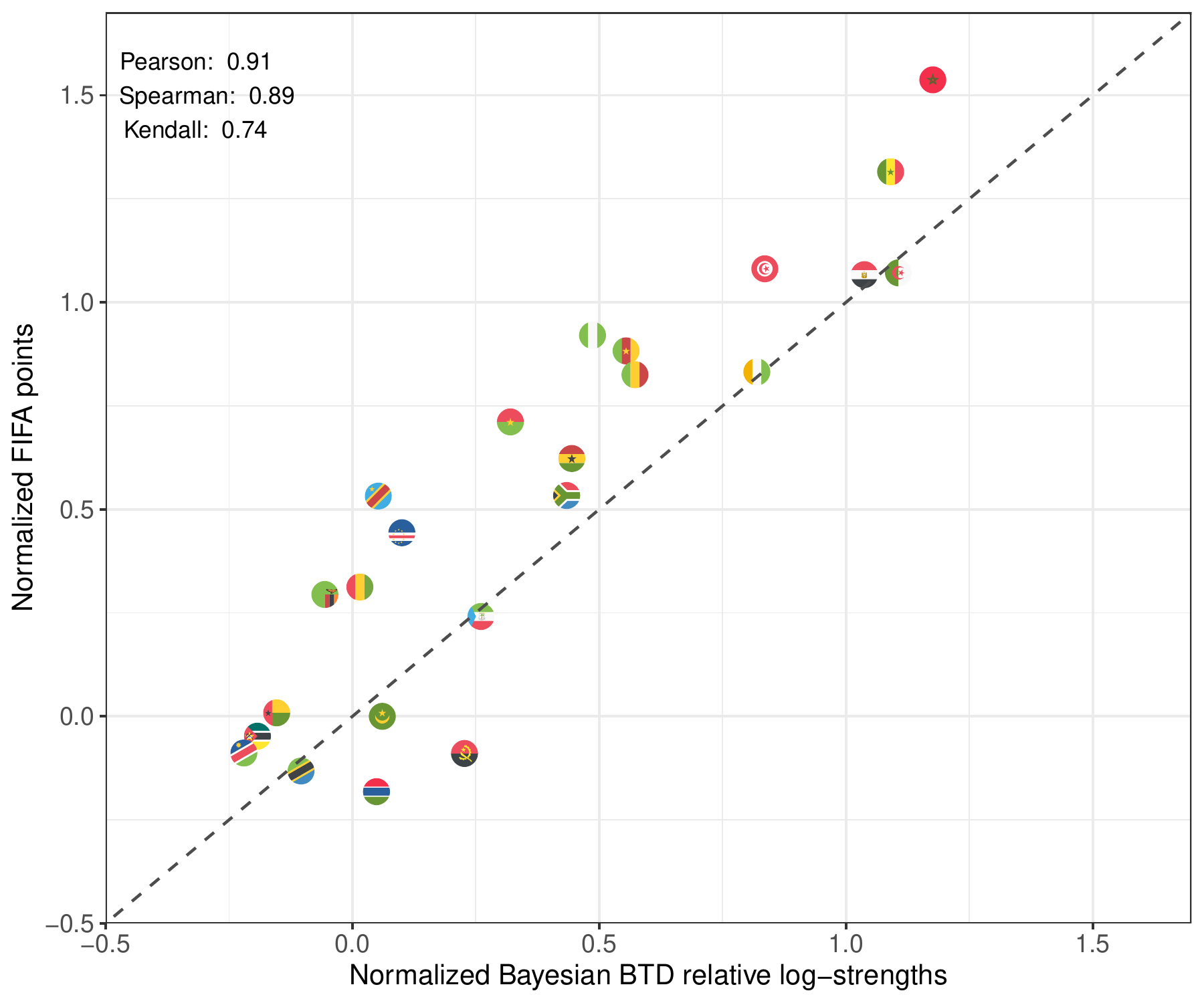}
\caption{2023 CAF Africa Cup of Nations. Scatterplot comparing FIFA ranking points and BTD relative log-strengths of teams in the Africa Cup of Nations.}
\label{ranks_comp_Africa_Cup}
\end{figure}

Figure \ref{ranks_comp_Africa_Cup} presents a scatterplot for the 2023 CAF Africa Cup of Nations, showing the relationship between normalized Bayesian BTD relative log-strengths and normalized FIFA points. The results are consistent with those from the 2022 World Cup, showing a Pearson correlation coefficient of $\rho_p = 0.91$, a Spearman correlation coefficient of $\rho_s = 0.89$, and a Kendall coefficient of $\tau = 0.74$, all indicating strong positive associations.

\begin{figure}[htb!]
    \centering   \includegraphics[width=14.5cm, height=14.5cm]{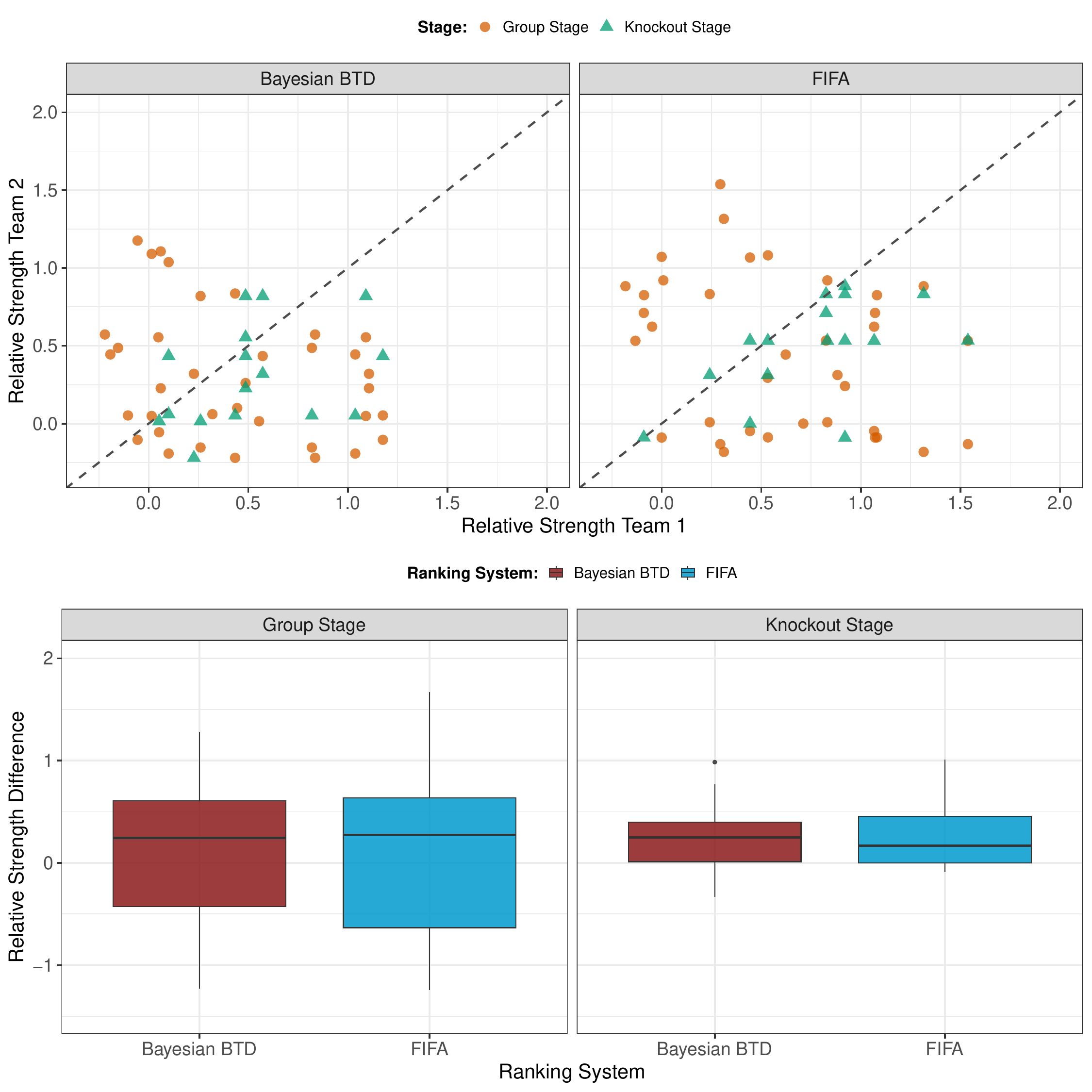}
        \caption{2023 CAF Africa Cup of
Nations. The top panel displays scatterplots comparing the normalized relative strengths of teams in both the group stage (orange dots) and knockout stage (green dots) under the two ranking systems. The dashed grey line represents the bisector. The bottom panel presents boxplots of the normalized relative strength differences for the FIFA ranking (blue) and the Bayesian BTD derived ranking (red) across the two Africa Cup of
Nations stages.}
        \label{ranks_Africa_Cup}
\end{figure}

The Bayesian BTD relative log-strengths exhibit less variability compared to the FIFA ranking points, in both the group and knockout stages, as illustrated in the top panel of Figure \ref{ranks_Africa_Cup}. Furthermore, the boxplots displayed in the bottom panel of Figure \ref{ranks_Africa_Cup} show a significant reduction in variance for both ranking systems as we move from the group stage to the knockout stage. However, contrary to what was observed in the World Cup, the Bayesian BTD relative log-strengths exhibit lower variance compared to the FIFA ranking points in the group stage. This may indicate that the Bayesian BTD relative log-strengths more accurately reflect the inherent similarity in team strengths within this tournament stage.

As reported in Table \ref{brier_Africa_Cup},  all the machine learning algorithms present similar prediction performance in the group stage of the AFCON. However, random forests show the weakest performance in the knockout stage. Furthermore, the other machine learning algorithms and statistical models exhibited similar levels of predictive accuracy in both stages, with statistical models performing slightly better in the group stage. In particular, the inclusion of this alternative ranking system generally improved the predictive performance of most of the statistical models evaluated in the knockout stage.
\begin{table}[t!]
\centering
\begin{tabular}{lcccc}
\toprule
\multirow{2}{*}{\textbf{Model}} & \multicolumn{2}{c}{\textbf{Group Stage}} & \multicolumn{2}{c}{\textbf{Knockout Stage}} \\ 
\cmidrule(lr){2-3} \cmidrule(lr){4-5} 
 & FIFA & BTD & FIFA & BTD \\ 
\midrule
Diag. Infl. & 0.679 & 0.682 & 0.681 & 0.677 \\ 
Biv. Pois. & 0.673 & 0.682 & 0.658 & 0.660 \\ 
Double Pois. & 0.670 & 0.677 & 0.670 & 0.656 \\ 
MARS & 0.679 & 0.690 & 0.645 & 0.650 \\ 
ANN & 0.703 & 0.702 & 0.666 & 0.661 \\ 
Random Forest & 0.736 & 0.687 & 0.834 & 0.884 \\ \bottomrule
\end{tabular}
\caption{2023 CAF Africa Cup of
Nations. Brier score for the FIFA ranking and the Bayesian BTD derived ranking across the two Africa Cup of
Nations stages.}
\label{brier_Africa_Cup}
\end{table}


\section{Discussion}\label{sec4}
This paper investigates the potential improvement in the predictive performance of statistical goal-based methods and machine learning result-based algorithms when a ranking system is incorporated as an additional predictor through its ranking points (relative strengths). We analyze data from the recent 2022 FIFA World Cup in Qatar and the 2023 CAF Africa Cup of Nations in Ivory Coast. Specifically, we explore the effectiveness of a Bayesian Bradley-Terry-Davidson derived ranking system in enhancing prediction accuracy compared to the well-established FIFA ranking system. We compare the performance of these two ranking systems across different tournament stages to identify their potential in predicting match outcomes.

While both the FIFA ranking points and the Bayesian BTD relative log-strengths provide valuable information for predicting outcomes, their effectiveness depends on the stage of the tournament. The FIFA ranking tends to be more accurate during the group stages of the 2022 World Cup, which features more heterogeneous team strengths. Conversely, the Bayesian BTD derived ranking is particularly effective in the group stage of the 2023 AFCON, and it also presents slightly better predictive performances in the knockout stages of both the World Cup and the AFCON. These stages typically exhibit smaller differences in team strengths. Consequently, this result suggests that the Bayesian BTD model effectively captures shifts in team strengths, making it especially valuable in tournaments where competing teams are similar, such as the AFCON, or in stages where the teams exhibit comparable strengths.

It is important to note that while the Bayesian BTD derived ranking is a valuable alternative, it is more computationally intensive compared to the standard FIFA ranking. Specifically, this approach involves a two-step procedure. First, the Bayesian BTD model needs to be computed to derive the teams' relative log-strengths, and only then their difference can be used as additional predictor in the models being considered.

However, the potential to enhance the accuracy and predictive performances of these models remains significant. Further development could involve refining the Bayesian BTD model to include additional variables that impact match outcomes. These could include the overall market value of the players involved in a specific team, the number of Champions League players, an indicator of the hosting country, or the teams in its neighborhood. Even economic variables such as the GDP per capita or the national population size may be interesting. Furthermore, the implementation of a dynamic methodology, that enables the continuous adjustment for fluctuations in team strength over the course of a season or tournament, could potentially result in enhanced prediction accuracy.

The potential application of Bayesian Bradley-Terry derived rankings, as an alternative for the FIFA ranking or similar systems, represents a promising area for research. Further comparative studies across different sports or competition structures will be conducted to validate the effectiveness of the Bradley-Terry models. As broadly remarked, the interplay between the type of competition, the adopted ranking, and the chosen methodology represents a hot topic for football modellers and deserves a deep and further understanding, both in international matches and domestic leagues.

\section*{Software and Data Availability}
All analyses were conducted in the R programming language version $4.2.3$ \citep{R_software_2023}. The code to reproduce this manuscript is openly available at \url{https://github.com/RoMaD-96/Bayesian_BTD}. The data are available on Kaggle at \url{https://www.kaggle.com/datasets/martj42/international-football-results-from-1872-to-2017}.

\section*{Acknowledgments}

This work has been supported by the project "SMARTsports: “Statistical Models and AlgoRiThms in sports. Applications in professional and amateur contexts, with able-bodied and disabled athletes”, funded by the MIUR Progetti di Ricerca di Rilevante Interesse Nazionale (PRIN) Bando 2022 - grant n. 2022R74PLE (CUP J53D23003860006).

\bibliographystyle{apalike}
\bibliography{main}

\end{document}